\documentclass[conference]{IEEEtran}
\IEEEoverridecommandlockouts
\usepackage{cite}
\usepackage{amsmath,amssymb,amsfonts}
\usepackage{algorithmic}
\usepackage{graphicx}
\usepackage{textcomp}
\usepackage{xcolor}
\usepackage{float}
\usepackage{tikz}
\usepackage{makecell}
\usepackage{multirow}
\usepackage{array}
\usepackage{academicons}
\definecolor{orcidlogocol}{HTML}{A6CE39}
\usepackage{booktabs}
\def\BibTeX{{\rm B\kern-.05em{\sc i\kern-.025em b}\kern-.08em
    T\kern-.1667em\lower.7ex\hbox{E}\kern-.125emX}}
\begin{document}

\title{Reliable Traffic Monitoring Using Low-Cost Doppler Radar Units\\
}

\author{\IEEEauthorblockN{Mishay Naidoo}
\IEEEauthorblockA{\textit{Radar Remote Sensing Group} \\
\textit{University of Cape Town}\\
Cape Town, South Africa \\
ORCID 0009-0007-8606-8485}
\and
\IEEEauthorblockN{Stephen Paine}
\IEEEauthorblockA{\textit{Radar Remote Sensing Group} \\
\textit{University of Cape Town}\\
Cape Town, South Africa \\
ORCID 0000-0001-8621-7005}
\and
\IEEEauthorblockN{Amit Kumar Mishra}
\IEEEauthorblockA{\textit{National Spectrum Centre} \\
\textit{Aberystwyth University}\\
Aberystwyth, UK \\
ORCID 0000-0001-6631-1539}
\and
\hspace{7.2cm}
\IEEEauthorblockN{Mohammed Yunus Abdul Gaffar}
\IEEEauthorblockA{\textit{\hspace{7.2cm}Radar Remote Sensing Group} \\
\hspace{7.2cm}\textit{University of Cape Town}\\
\hspace{7.2cm}Cape Town, South Africa \\
\hspace{7.2cm}ORCID 0000-0002-8877-4515}

}

\maketitle

\begin{abstract}
Road traffic monitoring typically involves the counting and recording of vehicles on public roads over extended periods. The data gathered from such monitoring provides useful information to municipal authorities in urban areas. This paper presents a low-cost, widely deployable sensing subsystem based on Continuous Wave Doppler radar. The proposed system can perform vehicle detection and speed estimation with a total cost of less than 100\,USD. The sensing system (including the hardware subsystem and the algorithms) is designed to be placed on the side of the road, allowing for easy deployment and serviceability. 
\end{abstract}

\section{Introduction}
Road traffic monitoring is key to infrastructure planning and typically involves counting the number of vehicles driving on a road at a given point in time, recording their velocities, and identifying the class of vehicle \cite{Bernas}. A traffic monitoring system should be capable of performing this task automatically, allowing the user to obtain traffic information easily and reliably. Such a system is useful in assisting municipal authorities by providing them with a better understanding of the number and class of vehicles travelling on their roads throughout the day. This information can help identify areas needing road upgrades, plan road maintenance around traffic, and provide insights into pollution emissions. A successful traffic monitoring system in a city requires several sensor nodes to be deployed in various locations which form part of a larger system of sensors. Having multiple sensor subsystems communicating to a central location allows the user to monitor many roads in a city simultaneously. Considering the financial cost of a multi-sensor system, the individual sensors should be low-cost.

Current sensing techniques for obtaining traffic data include pneumatic tubes \cite{Czyzewski}, video analysis \cite{Nemade}, Light Detection and Ranging (LiDAR) \cite{Czyzewski}, joint communication and sensing (JCAS) systems \cite{sardar2019vehicle, jana2023validation}, Frequency Modulated Continuous Wave (FMCW) radar \cite{Lim}, and Continuous Wave (CW) Doppler radar \cite{Czyzewski}. Pneumatic tubes are simple to implement and can identify both vehicle speeds and classify the vehicles based on width \cite{Czyzewski}. Two tubes are laid across the road each detecting when a vehicle passes over them. Using the time difference between each tube detection and the distance between each tube, the length and speed of the vehicle can be calculated \cite{Czyzewski}. Both video and FMCW radar-based solutions can monitor multiple lanes of traffic simultaneously and can accurately distinguish closely grouped vehicles \cite{Czyzewski, Lim}. LiDAR sensors emit a focused beam of coherent light and are less prone to measurement errors than radar \cite{Czyzewski}. However, existing systems have various limitations, examples include pneumatic tubes' inability to handle vehicles at high speeds \cite{Czyzewski}, video cameras offering poor performance in low visibility conditions \cite{Lim}, LiDAR being unable to easily obtain speed information from the detected targets \cite{Czyzewski}, and both LiDAR and FMCW radar modules being very costly. In contrast, CW Doppler radar is low-cost and can accurately measure vehicles at high speeds \cite{Fang} making it a suitable sensor for this application.

CW Doppler sensors have been used to successfully classify vehicles and robustly estimate the speed of vehicles \cite{Fang, Nguyen}. In these existing solutions, a node was fixed to the underside of a bridge or overpass, with the antenna beam directed downward toward the road. This setup was used to collect the data which was processed on a microcontroller to perform classification and speed estimation. Despite this, CW sensors are known to struggle in slow-moving traffic where vehicles come to a full stop \cite{Czyzewski} and are undetectable by the sensor.   

For a traffic monitoring system to cover a significant area of a city, many sensors must be deployed across different roads and the results need to be collated to provide effective monitoring. The existing sensors in \cite{Fang, Nguyen} are expensive due to manufacturing costs as well as the choice of Digital Signal Processor (DSP) devices used to sample and process the data. The cheapest of these nodes is approximately 200\,USD \cite{Fang}. Unfortunately, the high cost of these sensors makes manufacturing and deploying large quantities of them very expensive and impractical for wide-area traffic monitoring. Furthermore, the requirement for the sensor to be positioned on a bridge means that it can only be deployed on a limited number of roads, reducing the number of roads that can be monitored.

The original contribution of this paper is investigating the feasibility of using a low-cost CW Doppler radar to successfully count vehicle numbers and accurately estimate vehicle speeds in a widely deployable low-cost sensor system. The development of a low-cost Doppler radar system would benefit municipalities when expanding roads in the future and improving traffic flow. This investigation includes the design and testing of a prototype sensor solution using low-cost commercial off-the-shelf (COTS) components.

The rest of the paper is organized as follows: Section II describes the system design process, Section III shows the results followed by concluding remarks in Section IV.

\section{System Description}
\subsection{System Components}
\begin{figure}[H]
    \centering
    \includegraphics[width=0.45\textwidth]{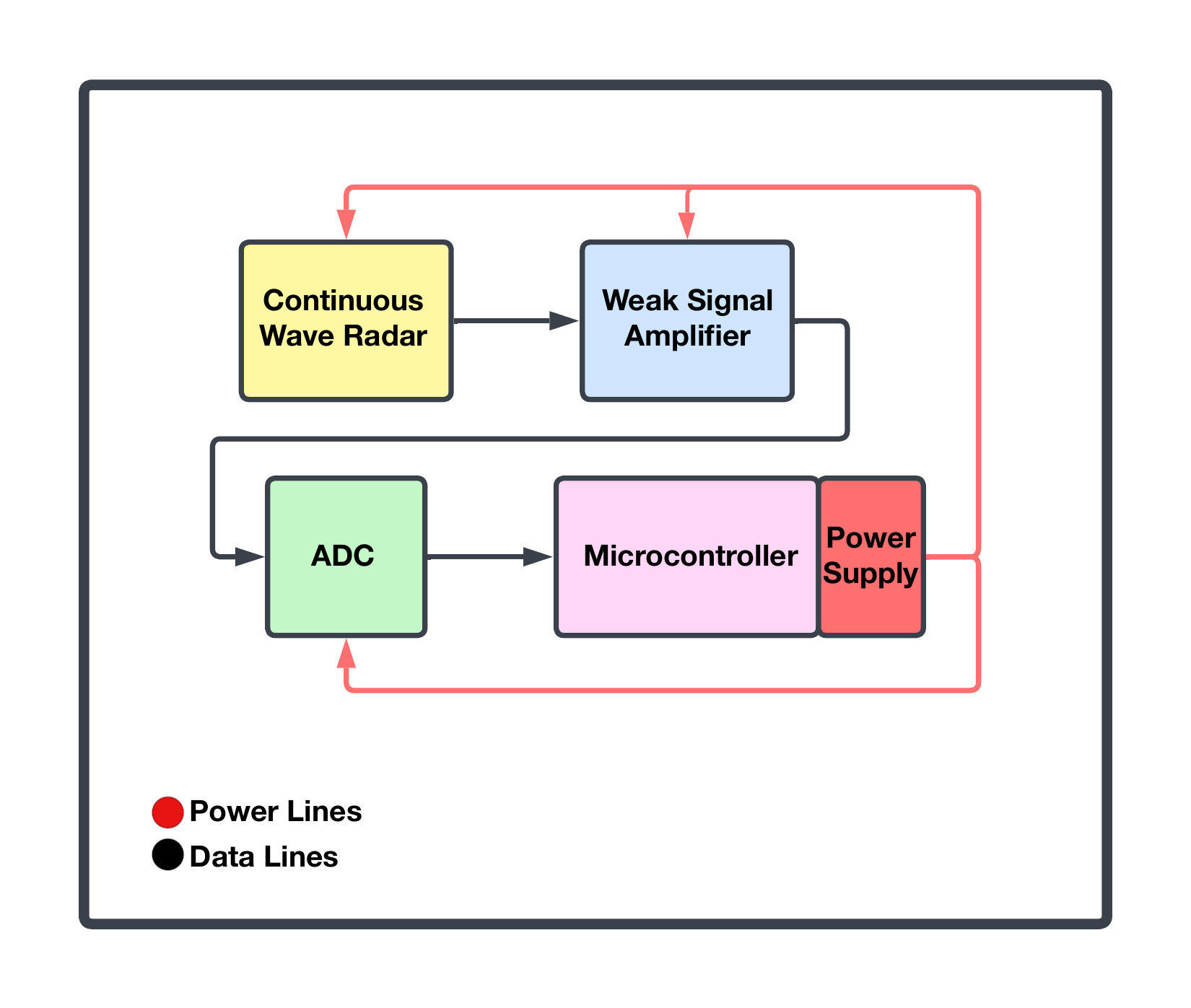}
    \caption{Component Interface Diagram}
    \label{fig:SystemBlockDiagram}
\end{figure}

The proposed traffic monitoring platform consists of five subsystems interfaced together for data collection and processing. These components included a CW radar, an amplifier to amplify the weak return of the Doppler sensor, an analog-to-digital converter (ADC) to sample the data, a microcontroller for data processing, and a power supply, as shown in Fig. \ref{fig:SystemBlockDiagram}.

The exact component used for each subsystem and their corresponding cost at the time of purchase is detailed in Table \ref{tab:Components}.

\begin{table}[H]
\centering
\caption{Component List}
\label{tab:Components}
\renewcommand{\arraystretch}{1.5} 
\begin{tabular}{|>{\centering\arraybackslash}m{3.5cm}|>{\centering\arraybackslash}m{3cm}|}
\hline
\textbf{Component Name} & \textbf{Cost [USD]} \\ \hline
CDM324/IPM-165           & 6.47              \\ \hline
LM358 Weak-Signal Amplifier & 1.86           \\ \hline
Soundblaster G3          & 68                \\ \hline
Raspberry Pi Zero 2 W    & 15                \\ \hline
\multicolumn{1}{|r|}{\textbf{Total}} & \textbf{91.33} \\ \hline
\end{tabular}
\end{table}

The \textit{CDM324/IPM-165} by \textit{InnoSenT} is an affordable K-band CW radar module that operates at 24.125\,GHz and is used as the core sensing component within the system. The relationship between the Doppler shift and a target's radial velocity for this particular sensor is given by Equation (\ref{eqn: Doppler Frequency}) \cite{HB100Datasheet}:
\begin{equation}
\label{eqn: Doppler Frequency}
    F_{Doppler}\,[Hz] = 44.68 \times Velocity\,[km/h]
\end{equation}
This sensor outputs the beat frequency created by moving targets in the antenna beam. This output is very low power, due to the distance between the target and the sensor's receive antenna, and has a peak-to-peak voltage in the milli-volt range.  An \textit{LM358} weak-signal amplifier module with a maximum linear gain of 100 was used to amplify the radar's output signal. 

After amplification, the analog signal was sampled using the \textit{Soundblaster G3} soundcard by \textit{Creative Labs}. The output frequencies of the Doppler sensor are in the audible range (20\,kHz or less) making the audio soundcard a suitable sampling tool for this application. The soundcard ADC has a resolution of 16 bits and a sampling rate of 48\,ksps. Equation \eqref{eqn: Doppler Frequency} demonstrates that a sampling rate of 48\,ksps can accurately sample velocities up to 537\,km/h according to the Nyquist frequency of the soundcard. This soundcard was the most expensive component in the system priced at 68\,USD, and will be replaced with a cheaper solution in future iterations to reduce the total bill of materials (BOM) to below 50\,USD.

The sound card was intended to interface with a \textit{Raspberry Pi Zero 2 W}, and all the algorithms developed were designed for execution on this single-board computer (SBC). Priced at 15\,USD, the Raspberry Pi Zero 2 W is a budget-friendly option that seamlessly interfaces with the USB soundcard for data retrieval. However, for the initial phase of this work, testing was done on a laptop.

\subsection{Sensor Placement}
One of the requirements of the traffic monitoring system was the wide deployability of the sensing solution. This included avoiding the need to place the sensing subsystem on a bridge or overpass above the road. Instead, this system was designed to be placed on the side of the lane being monitored. The sensor was placed at an angle of approximately $20^\circ$ to the axis of the road such that vehicles would be traveling almost directly toward or away from the antenna boresight, resulting in a large Doppler shift.

\begin{figure}[H]
    \centering
    \includegraphics[width=0.35\textwidth, angle=90]{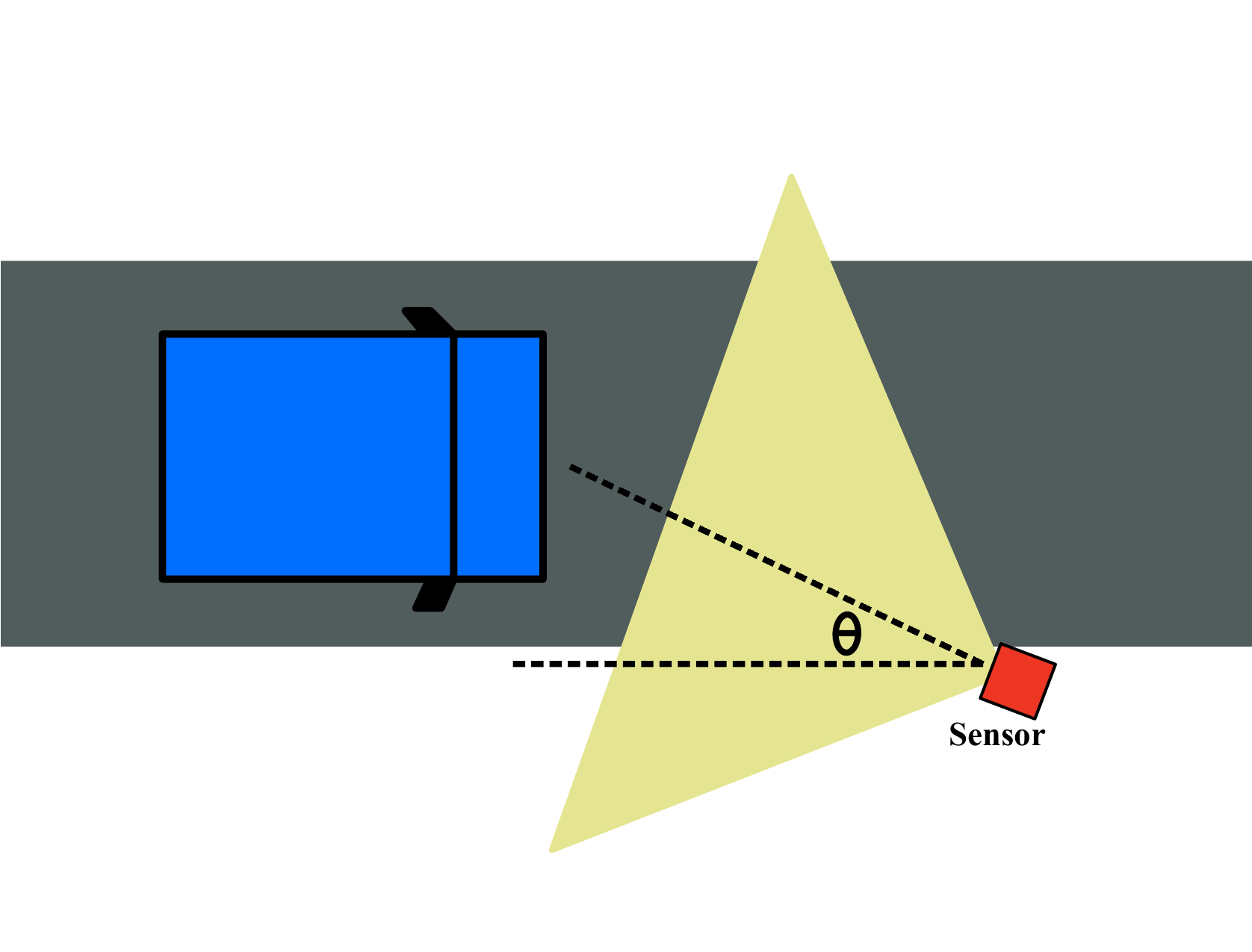}
    \caption{Sensor System Placement Top-Down View}
    \label{fig:SensorPlacement}
\end{figure}

\begin{figure}[H]
    \centering
    \includegraphics[width=0.35\textwidth, angle=90]{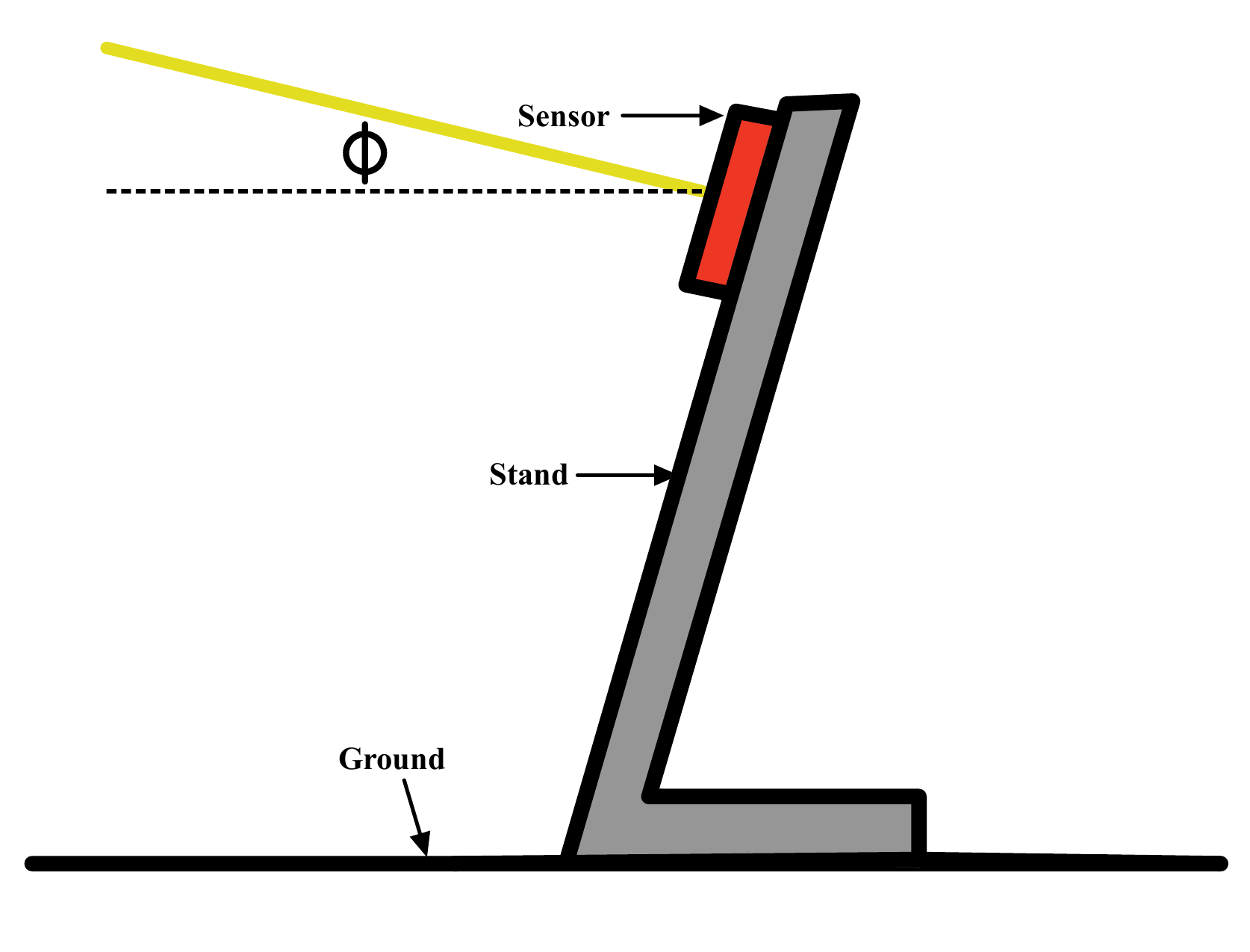}
    \caption{Sensor System Placement Side View}
    \label{fig:SensorPlacement2}
\end{figure}

This design allows the user to deploy a sensor node onto traffic lights or roadside lamps that are present on most roads in urban areas. The system was designed to work close to the ground with a low grazing angle where the antenna beam is angled up toward the approaching vehicles as seen in Fig. \ref{fig:SensorPlacement2}. While this approach ensures easy sensor deployment, it limits the system to single-lane monitoring of traffic per sensor node.

\subsection{Data Processing}
\begin{figure}[H]
    \centering
    \begin{tikzpicture}
        \node[anchor=south west,inner sep=0] at (0,0) {\includegraphics[width=0.2\textwidth, angle = 90]{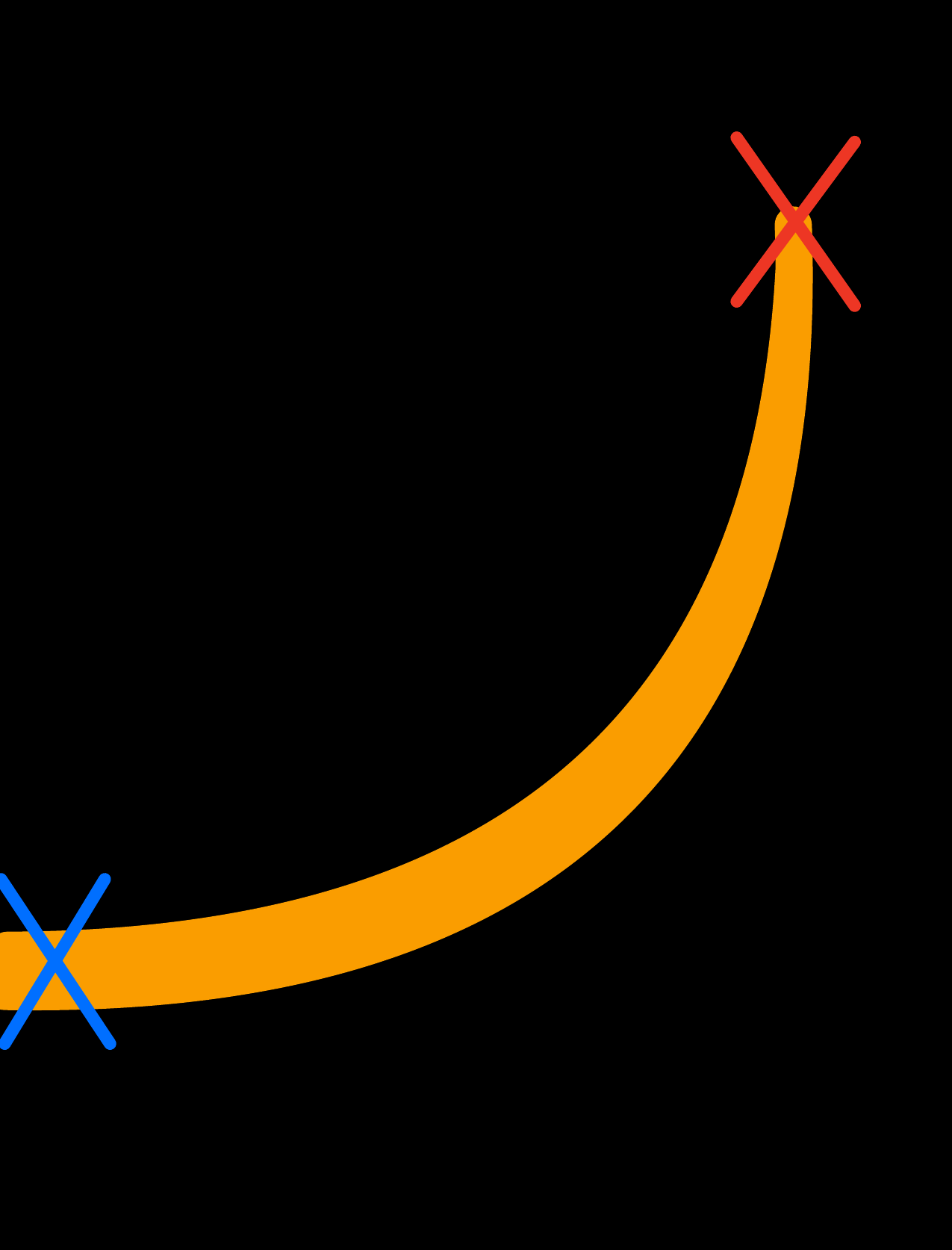}};

        \node[below=0.3cm, align=center] at (current bounding box.south) {Time [s]};

        \node[rotate=90, above=0.3cm, align=center, xshift = 0.4cm] at (current bounding box.west) {Frequency [Hz]};
    \end{tikzpicture}
    
    \caption{Example Hand-Drawn Illustration of Expected Spectrogram Created by Vehicle Driving Towards Sensor}
    \label{fig:SpectrogramIllustration}
\end{figure}

The data obtained from the soundcard was stored as a Waveform Audio File (WAV) file. A Python script was used to visualize and process the data. The primary analysis tool used to visualize the data was the spectrogram. The spectrogram is a frequency-time heatmap and showcases the frequencies apparent in a signal at any given time during the recording. This plot is generated by performing multiple Short-Time Fourier Transforms (STFT) on windowed sections of data \cite{Fang}. The spectrograms displayed in this paper were created using a Hanning window with a size of 4096 samples and an overlap of 512 samples. The position, time of detection, and velocity of each target could be seen on the spectrogram of the captured data as illustrated in Fig. \ref{fig:SpectrogramIllustration}.

Fig. \ref{fig:SpectrogramIllustration} shows an illustration of the expected spectrogram created by a vehicle driving towards the sensor when placed in the configuration shown in Fig. \ref{fig:SensorPlacement}. The y-axis shows the Doppler shift created by the vehicle and the x-axis shows the time of occurrence. The Doppler shift decreases with time as the vehicle gets closer to the sensor because the CW radar measures the radial velocity of the vehicle. From the sensor placement in Fig. \ref{fig:SensorPlacement} it can be observed that the angle between the vehicle velocity vector and the antenna beam increases until a maximum of $90^\circ$ when the vehicle is directly beside the sensor. This corresponds to a Doppler shift that approaches zero and is marked by a blue 'X` in Fig. \ref{fig:SpectrogramIllustration}. Therefore, the largest Doppler frequency apparent in the spectrogram is the closest frequency related to the target's actual velocity, indicated by the red 'X` in Fig. \ref{fig:SpectrogramIllustration}. However, because the sensor is situated at an angle to the road (this angle was selected to be $20^\circ$) the maximum frequency obtained from the spectrogram was multiplied by an additional factor when converting to velocity. Additionally, because the sensor was placed on the ground and angled upwards, this angle had to be accounted for (this angle was also $20^\circ$). Equation \eqref{eqn: Doppler Frequency} can be updated to include these factors:
\begin{equation}
\label{eqn:Updated Doppler Frequency}
    F_{Doppler}\,[Hz] = 44.68 \times V_{max}\,[km/h] \times cos(\theta) \times cos(\phi)
\end{equation}
Here, $\theta$ is the angle between the antenna beam and the vehicle's velocity vector (seen in Fig. \ref{fig:SensorPlacement}) and $\phi$ is the angle between the antenna beam and the ground (seen in Fig. \ref{fig:SensorPlacement2}). One challenge that needed to be overcome was the system's susceptibility to noise from the power supply and low-frequency noise from the amplifier. To filter the noise and weak interference signals, a power threshold was applied to the data, removing all data points with a weak signal-to-noise ratio (SNR) from the spectrogram. This threshold also filtered out vehicle(s) in adjacent lanes to the one being monitored as these vehicle(s) returned a weaker signal due to the increased distance between the sensor and the vehicle(s). Furthermore, the low-frequency noise was removed by discarding all data points below a threshold frequency. Fortunately, the higher power noise was below 700\,Hz which corresponds to a velocity of 15\,km/h or less. Future iterations of the sensor system will include improved hardware-based filtering to overcome these issues before they leak into the data.

The next step in the algorithm extracted the highest frequency created by each target in the thresholded spectrogram and the corresponding time each target appeared in the plot. The frequency was converted to a velocity using Equation \eqref{eqn:Updated Doppler Frequency} and the resultant output was stored on the processing device.

\section{Results}
\begin{figure}[H]
    \centering
    \includegraphics[width=0.25\textwidth, angle=90]{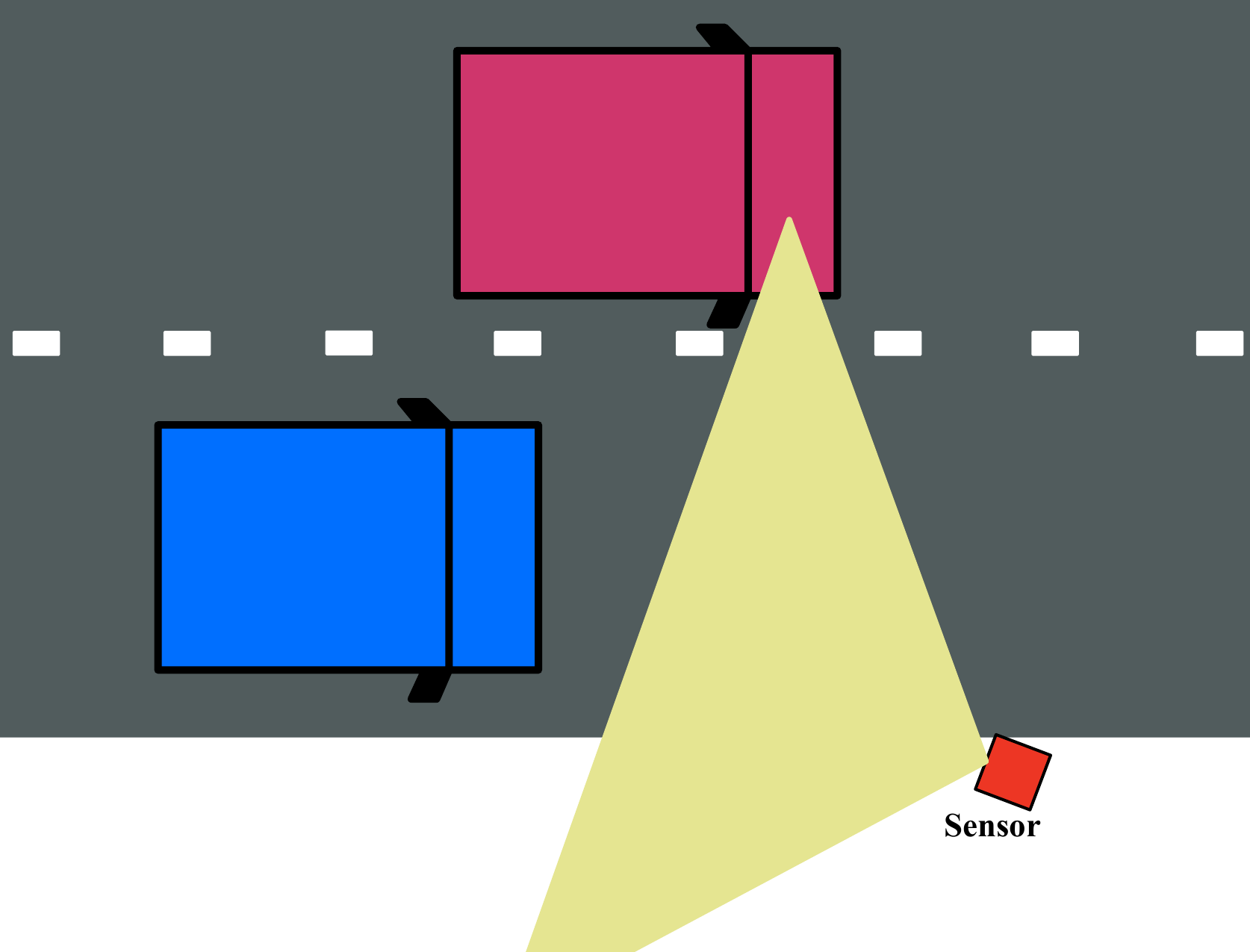}
    \caption{Two Lane Test Visualisation}
    \label{fig:TwoLaneSetup}
\end{figure}

This system was tested on a two-lane road with a speed limit of 60\,km/h. The device was placed as close as possible to the traffic (approximately 10\,cm away from the road). Fig. \ref{fig:TwoLaneSetup} illustrates the system set-up in the conducted tests.

\begin{figure}[H]
    \centering
    \begin{tikzpicture}
        \node[anchor=south west,inner sep=0] at (0,0) {\includegraphics[width=0.45\textwidth]{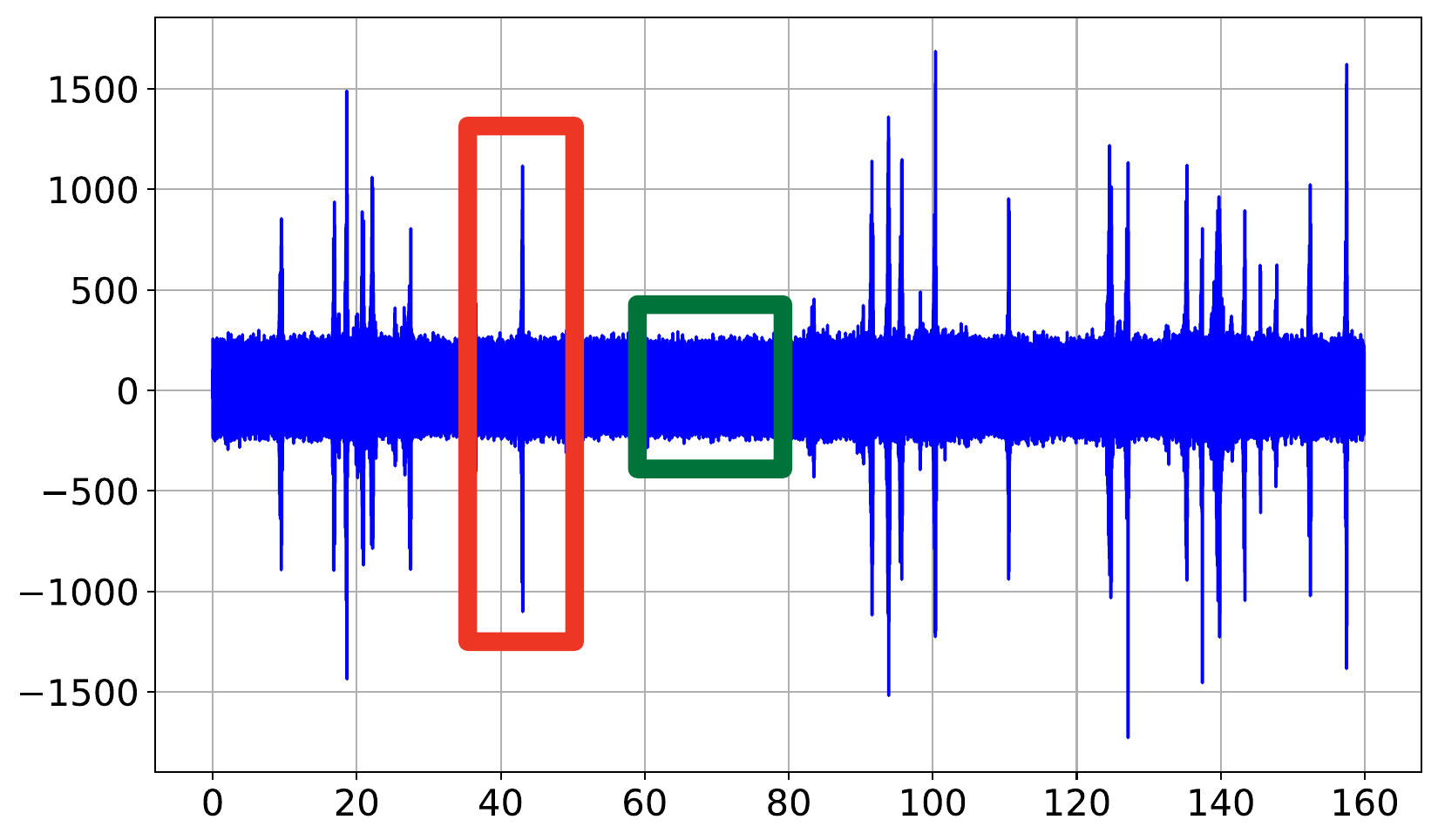}};

        \node[below=0cm, align=center, xshift = 0.3cm] at (current bounding box.south) {Time [s]};

        \node[rotate=90, above=0cm, align=center, xshift = 0.3cm] at (current bounding box.west) {Magnitude};
    \end{tikzpicture}
    
    \caption{Time Domain Waveform}
    \label{fig:TimeDomainResult}
\end{figure}

\begin{figure}[H]
    \centering
    \begin{tikzpicture}
        \node[anchor=south west,inner sep=0] at (0,0) {\includegraphics[width=0.45\textwidth]{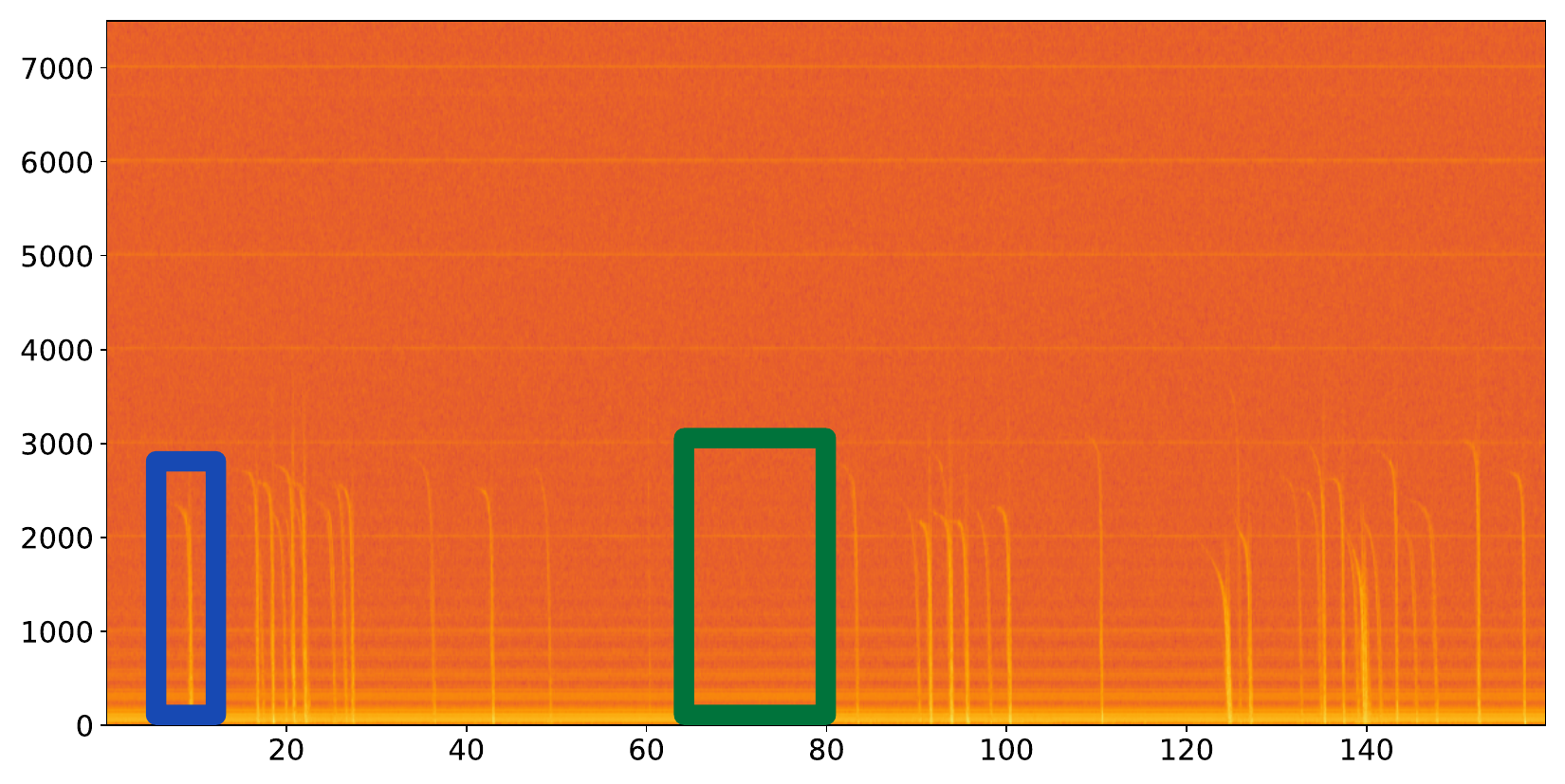}};

        \node[below=0cm, align=center, xshift = 0.2cm] at (current bounding box.south) {Time [s]};

        \node[rotate=90, above=0cm, align=center, xshift = 0.3cm] at (current bounding box.west) {Frequency [Hz]};
    \end{tikzpicture}
    
    \caption{Spectrogram of Data in Fig. \ref{fig:TimeDomainResult}}
    \label{fig:SpectrogramResult}
\end{figure}

\begin{figure}[H]
    \centering
    \begin{tikzpicture}
        \node[anchor=south west,inner sep=0] at (0,0) {\includegraphics[width=0.45\textwidth]{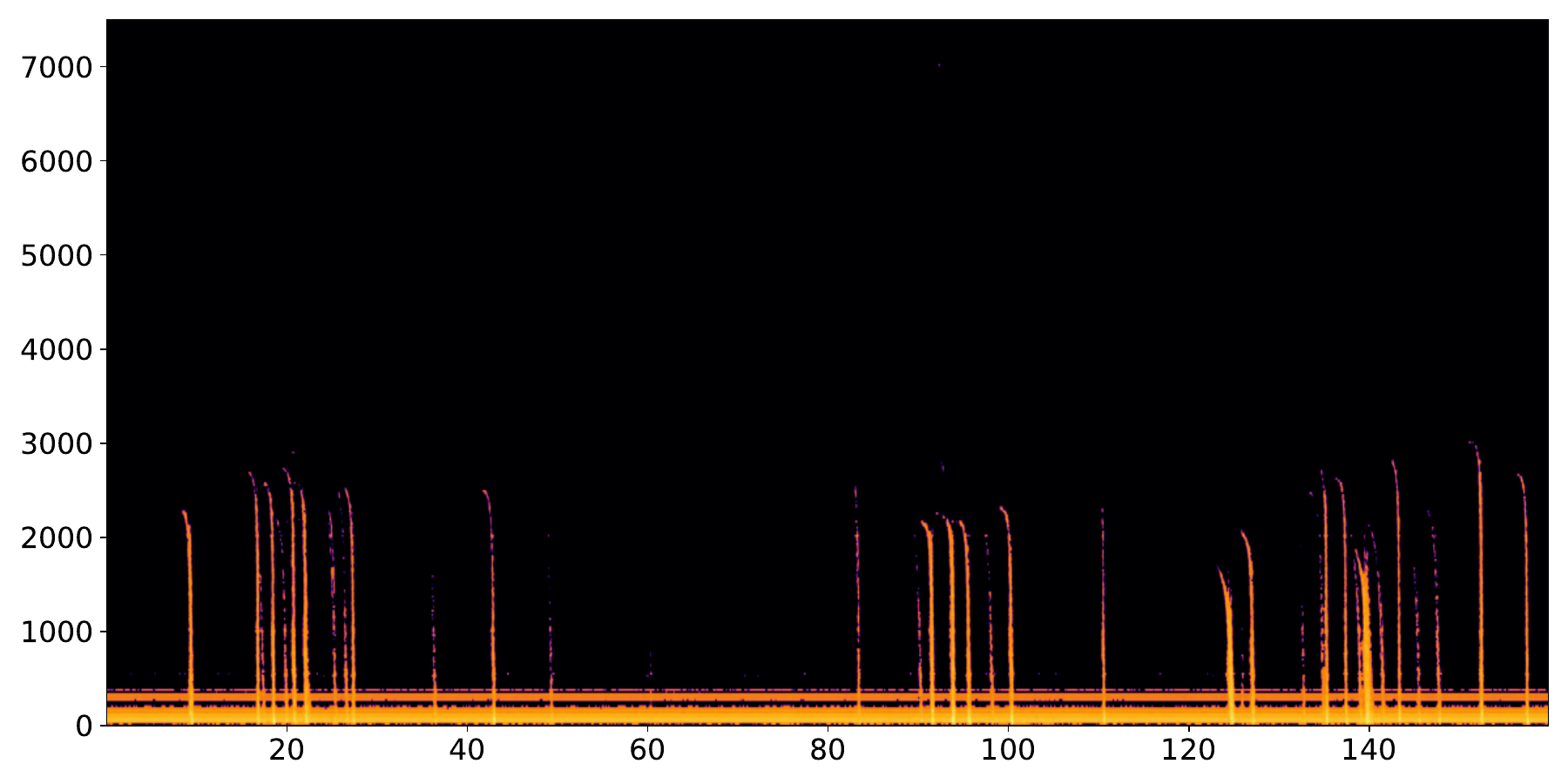}};

        \node[below=0cm, align=center, xshift = 0.2cm] at (current bounding box.south) {Time [s]};

        \node[rotate=90, above=0cm, align=center, xshift = 0.3cm] at (current bounding box.west) {Frequency [Hz]};
    \end{tikzpicture}
    
    \caption{Thresholded Spectrogram}
    \label{fig:SpectrogramThresholdedResult}
\end{figure}

\begin{figure}[H]
    \centering
    \begin{tikzpicture}
        \node[anchor=south west,inner sep=0] at (0,0) {\includegraphics[width=0.45\textwidth]{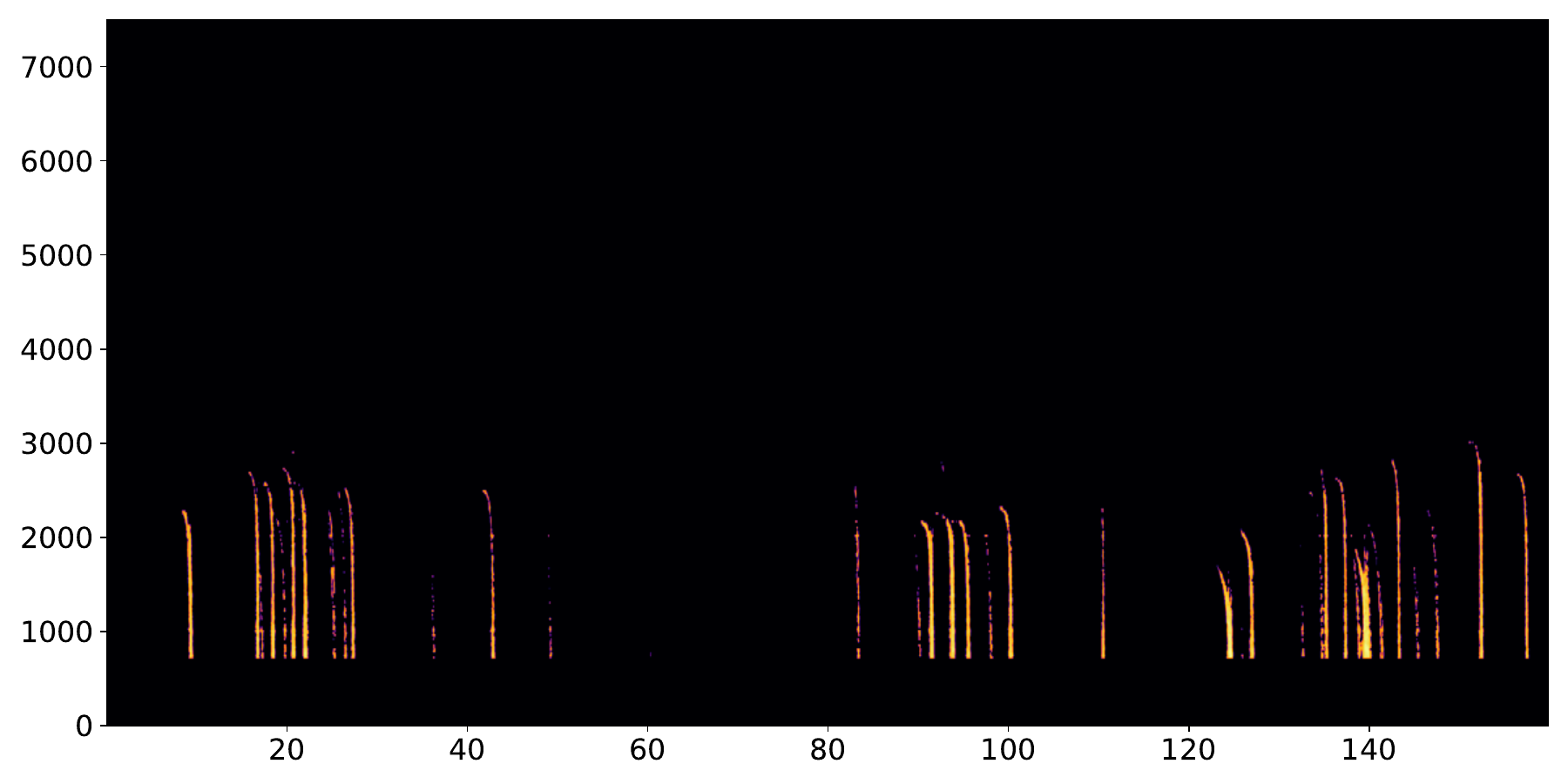}};

        \node[below=0cm, align=center, xshift = 0.2cm] at (current bounding box.south) {Time [s]};

        \node[rotate=90, above=0cm, align=center, xshift = 0.3cm] at (current bounding box.west) {Frequency [Hz]};
    \end{tikzpicture}
    
    \caption{Thesholded Spectrogram with Low-Frequencies Removed}
    \label{fig:SpectrogramLowFreqResult}
\end{figure}

\begin{figure}[H]
    \centering
    \begin{tikzpicture}
        \node[anchor=south west,inner sep=0] at (0,0) {\includegraphics[width=0.45\textwidth]{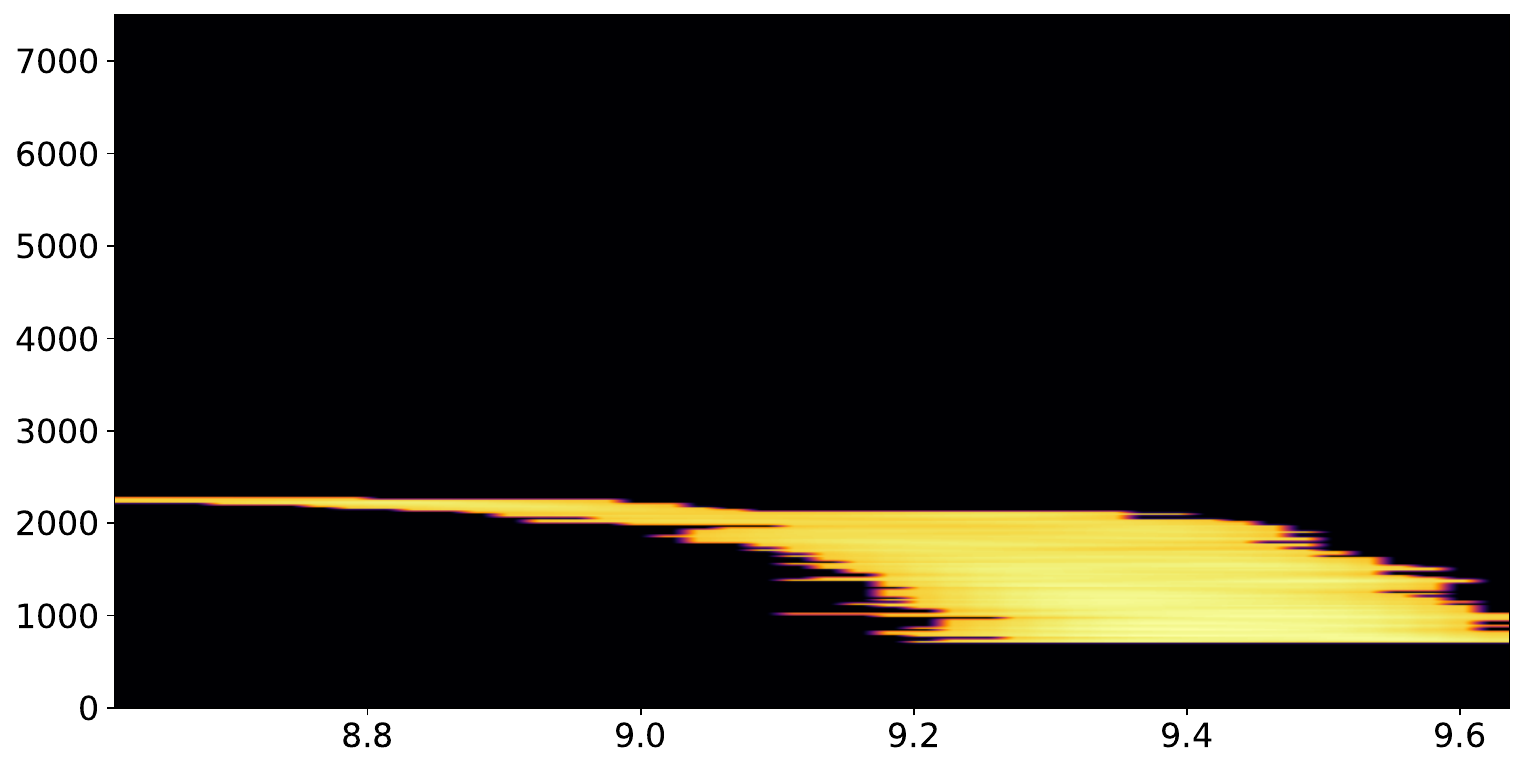}};

        \node[below=0cm, align=center, xshift = 0.2cm] at (current bounding box.south) {Time [s]};

        \node[rotate=90, above=0cm, align=center, xshift = 0.3cm] at (current bounding box.west) {Frequency [Hz]};
    \end{tikzpicture}
    
    \caption{Isolated Target Spectrogram}
    \label{fig:SpectrogramIsolatedResult}
\end{figure}

Data was collected by the system containing vehicle(s) moving towards the sensor and later post-processed to extract the target information. Fig. \ref{fig:TimeDomainResult} to Fig. \ref{fig:SpectrogramIsolatedResult} illustrate the plots generated in each stage of processing. This includes a plot of the raw amplified data sampled by the soundcard as well as spectrogram plots of the data in each stage of filtering. Lastly, an isolated target spectrogram after all processing is displayed.

The time domain representation of the data seen in Fig. \ref{fig:TimeDomainResult} reveals little information on individual targets and is not the ideal method of data representation for this investigation. Each target can be identified in the plot by the spike in magnitude highlighted by the red box in Fig. \ref{fig:TimeDomainResult}, however, no speed information can be gathered from this plot. This plot clearly shows a large amount of noise in the waveform which has been highlighted by the green box in Fig. \ref{fig:TimeDomainResult}. The individual targets appear much clearer in the spectrogram representation of the data in Fig. \ref{fig:SpectrogramResult} however, the SNR is very weak due to the noise within the system as discussed. The return from a single vehicle has been highlighted by the blue box in Fig. \ref{fig:SpectrogramResult}.

The background noise, highlighted by the green box in Fig. \ref{fig:SpectrogramResult}, was effectively removed by the thresholding applied to the spectrogram as seen in Fig. \ref{fig:SpectrogramThresholdedResult}.

 The high-power, low-frequency noise is visible in this thresholded plot but was successfully removed by discarding the low-frequency data as seen in Fig. \ref{fig:SpectrogramLowFreqResult}. 

\begin{figure}[H]
    \centering
    \begin{tikzpicture}
        \node[anchor=south west,inner sep=0] at (0,0) {\includegraphics[width=0.45\textwidth]{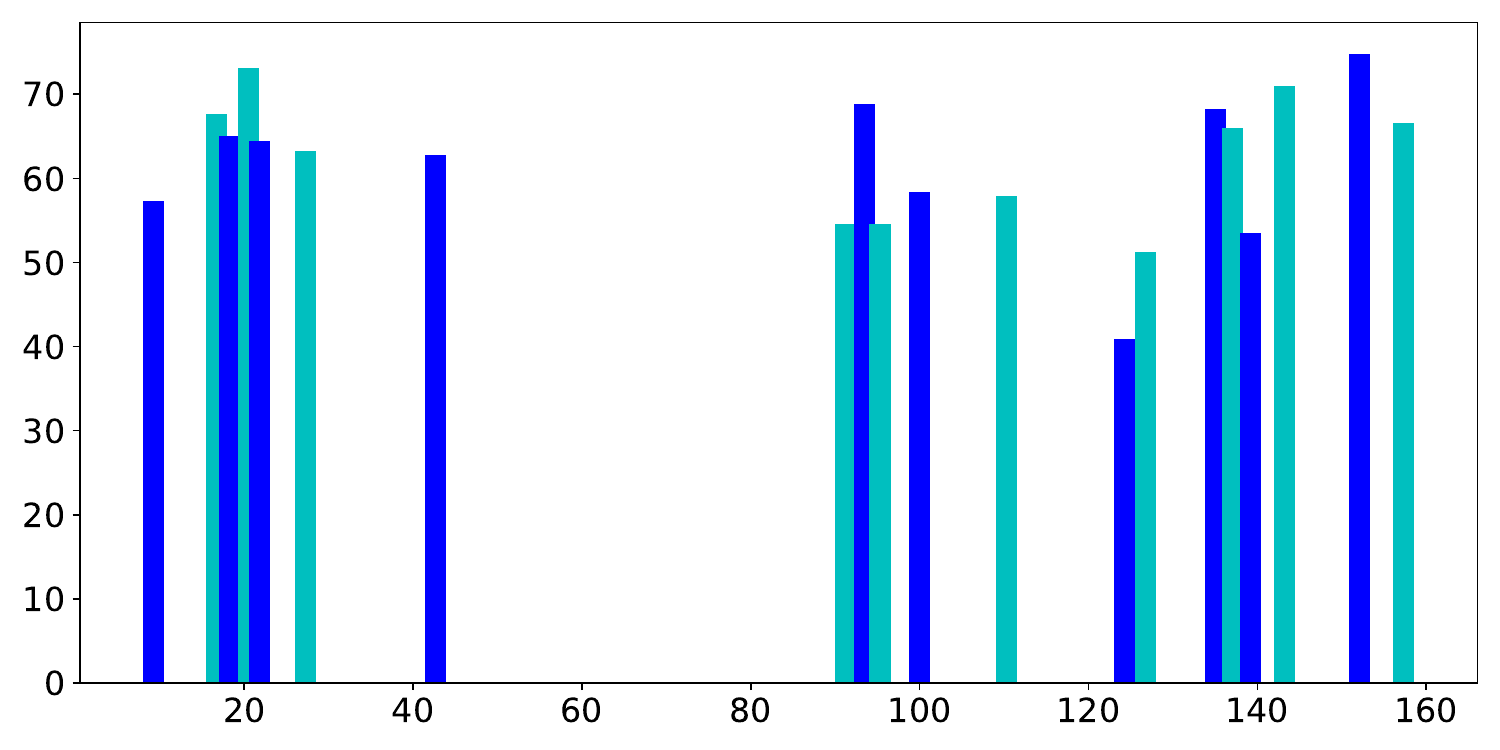}};

        \node[below=0cm, align=center, xshift = 0cm] at (current bounding box.south) {Time [s]};

        \node[rotate=90, above=0cm, align=center, xshift = 0.3cm] at (current bounding box.west) {Speed [km/h]};
    \end{tikzpicture}
    
    \caption{Identified Targets and Target Velocities (Note: The two different shades of blue are used to differentiate closely grouped targets)}
    \label{fig:BarGraphResults}
\end{figure}

Fig. \ref{fig:SpectrogramIsolatedResult} shows a single target isolated from the filtered plot in Fig. \ref{fig:SpectrogramLowFreqResult}. The shape of the measured spectrogram in Fig. \ref{fig:SpectrogramIsolatedResult} agrees with the expected plot illustrated in Fig. \ref{fig:SpectrogramIllustration}. The highest Doppler frequency created by the target was extracted and found to be 2261.1\,Hz. This was used in Equation (\ref{eqn:Updated Doppler Frequency}) to estimate the vehicle's speed resulting in a velocity estimate of 57.3\,km/h. Each target was isolated from the spectrogram and velocities were extracted.  Fig.\,\ref{fig:BarGraphResults} shows each detected vehicle, time of detection, and respective velocity. 

When conducting the tests, a recording of every car that passed the sensor in both the close and far lanes was made manually by the researcher conducting the test. These recordings were used as ground truth data to compare with the sensor data.
Table \ref{tab:VehicleCount} shows that in all the tests conducted, every vehicle in the far lane was successfully filtered out by the thresholding. Only a single car in the close lane passed the sensor unidentified.  

\begin{table}[H]
\caption{Number of Vehicles in Conducted Tests}
\renewcommand{\arraystretch}{1.2} 
\setlength{\tabcolsep}{10pt} 
\centering
\begin{tabular}{|l|l|l|l|}
\hline
\textbf{\begin{tabular}[c]{@{}l@{}}Number of\\ Cars in Close\\ Lane\end{tabular}} & \textbf{\begin{tabular}[c]{@{}l@{}}Number of\\ Cars in Far\\ Lane\end{tabular}} & \textbf{\begin{tabular}[c]{@{}l@{}}Number of\\ Cars in \\ Close Lane \\ Identified\end{tabular}} & \textbf{\begin{tabular}[c]{@{}l@{}}Number of\\ Cars in Far\\ Lane\\ Filtered Out\end{tabular}} \\ \hline
28                                                                                & 24                                                                              & 27                                                                                               & 24                                                                                              \\ \hline
\end{tabular}
\label{tab:VehicleCount}
\end{table}

\begin{figure}[H]
    \centering
    \begin{tikzpicture}
        \node[anchor=south west,inner sep=0] at (0,0) {\includegraphics[width=0.45\textwidth]{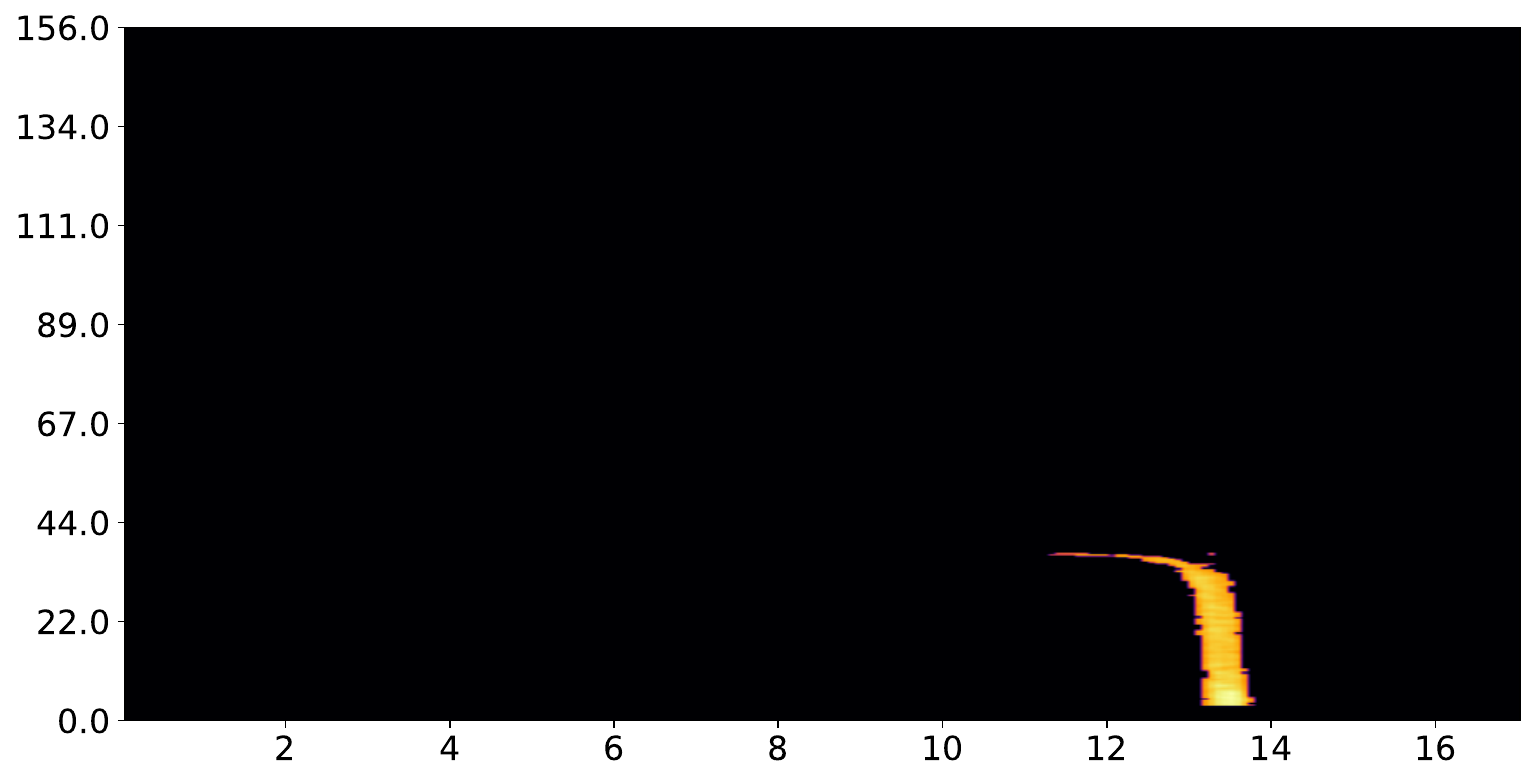}};

        \node[below=0cm, align=center, xshift = 0.2cm] at (current bounding box.south) {Time [s]};

        \node[rotate=90, above=0cm, align=center, xshift = 0.3cm] at (current bounding box.west) {Velocity [km/h]};
    \end{tikzpicture}
    
    \caption{Controlled Test Spectrogram}
    \label{fig:ControlledTestResult}
\end{figure}

Additionally, an isolated test with the same sensor configuration, but on a quieter single-lane road, was done with a vehicle traveling at a known velocity. In this test, a vehicle was driven towards the sensor at 40\,km/h, measured by the vehicle's speedometer. Fig. \ref{fig:ControlledTestResult} illustrates the processed spectrogram obtained from this controlled test. The velocity extracted from this plot was 42.3 km/h. 

The controlled test displayed a discrepancy of 2.3\,km/h, with the system recording a speed of 42.3\,km/h and the vehicle's speedometer displaying 40\,km/h. This controlled test was repeated a second time and once again the recorded velocity was 42.3\,km/h. These tests both displayed an error of $\frac{2.3}{40} \times 100 = 5.75\%$. The obtained velocities before the additional angular factors, as shown in Equation (\ref{eqn:Updated Doppler Frequency}), were both 37.4\,km/h. The vehicle velocity perceived by the radar was less than the vehicle's actual velocity which is expected due to the angle between the velocity vector and antenna beam.

The first tests prove that the sensor system can record targets driving past the sensor, with only a single target passing the sensor undetected. This indicates that the configuration with the sensor on the side of the road is feasible for implementation. Furthermore, the controlled tests revealed that the sensor system can accurately capture speed information when placed in this configuration. 

The error seen in the controlled tests can be attributed to either inaccuracies in the angle correction factors or a CW sensor accuracy limitation. The consistent error seen in both controlled tests indicates that it is most likely a correction factor error which can be easily updated. The source of the error can be determined by repeating tests with a third mechanism of speed capture (such as GPS) to compare with the speedometer and radar readings. 

The single vehicle that passed by the sensor undetected in the first test was travelling at 93\,km/h which was significantly faster than the other vehicles observed in the test. The speed of the vehicle resulted in a weaker SNR in the spectrogram and the target was filtered out by the thresholding. The target was visible in the unfiltered spectrogram and its speed was extracted from that plot. This highlights the need for improvement in the filtration algorithm or a potential high-speed sensing limitation.

\section{Conclusion and Future Work}
In this paper, the effectiveness of a low-cost sensing system using Doppler-only CW radar in a traffic monitoring context is presented. A cost-effective system was developed, priced at less than 100 USD. The system was able to monitor a single lane of traffic accurately, on multi-lane roads, with speed limits of 60\,km/h (this was the only speed limit tested and is not necessarily a system limitation). Furthermore, the system functioned when positioned on the side of the road allowing for easy deployment of each sensor in an urban area. This initial work has successfully demonstrated that a low-cost system has the potential for use in traffic monitoring and with some refinement and additional testing, could be deployed throughout cities to provide important traffic information.

\subsection{Future Work}
To further develop the traffic monitoring system used in this paper, more tests must be conducted. The efficacy of the system was only tested on a road with a speed limit of 60\,km/h and must still be tested in high-speed scenarios such as highways with speeds of up to 120\,km/h. Furthermore, the system was never tested in extremely dense traffic scenarios where traffic is stopping and restarting in front of the sensor. 

The cost of the system can still be significantly reduced by using an alternative ADC to the soundcard. The \textit{Soundblaster G3} attributed to 74\% of the system's total cost. The soundcard has additional audio processing capabilities that were not required in this application and can be replaced by a standalone ADC. High-quality 16bit ADCs can be purchased for as low as 8\,USD.

Furthermore, this paper investigated the development of an individual node in a large system of sensors. An investigation into the communication between nodes throughout a city must be conducted. These nodes could interface with a communication protocol such as LoRaWAN (Long Range Wide Area Network) allowing all nodes to remain connected across large areas \cite{Devalal}. 

Lastly, an investigation into the feasibility of vehicle classification from the data obtained from the system should be conducted. Fang et al. \cite{Fang} achieved vehicle classification using a similar system and these methods could be applied to this data to increase the amount of useful information obtained.

\bibliographystyle{IEEEtran}
\bibliography{Bibliography} 

\begin{thebibliography}{10}
\providecommand{\url}[1]{#1}
\csname url@samestyle\endcsname
\providecommand{\newblock}{\relax}
\providecommand{\bibinfo}[2]{#2}
\providecommand{\BIBentrySTDinterwordspacing}{\spaceskip=0pt\relax}
\providecommand{\BIBentryALTinterwordstretchfactor}{4}
\providecommand{\BIBentryALTinterwordspacing}{\spaceskip=\fontdimen2\font plus
\BIBentryALTinterwordstretchfactor\fontdimen3\font minus \fontdimen4\font\relax}
\providecommand{\BIBforeignlanguage}[2]{{%
\expandafter\ifx\csname l@#1\endcsname\relax
\typeout{** WARNING: IEEEtran.bst: No hyphenation pattern has been}%
\typeout{** loaded for the language `#1'. Using the pattern for}%
\typeout{** the default language instead.}%
\else
\language=\csname l@#1\endcsname
\fi
#2}}
\providecommand{\BIBdecl}{\relax}
\BIBdecl

\bibitem{Bernas}
\BIBentryALTinterwordspacing
M.~Bernas, B.~Płaczek, W.~Korski, P.~Loska, J.~Smyła, and P.~Szymała, ``A survey and comparison of low-cost sensing technologies for road traffic monitoring,'' \emph{Sensors}, vol.~18, no.~10, 2018. [Online]. Available: \url{https://www.mdpi.com/1424-8220/18/10/3243}
\BIBentrySTDinterwordspacing

\bibitem{Czyzewski}
A.~Czyżewski, S.~Cygert, G.~Szwoch, J.~Kotus, D.~Weber, M.~Szczodrak, D.~Koszewski, K.~Jamroz, W.~Kustra, A.~Sroczyński, T.~Śmiałkowski, and P.~Hoffmann, ``Comparative study on the effectiveness of various types of road traffic intensity detectors,'' in \emph{2019 6th International Conference on Models and Technologies for Intelligent Transportation Systems (MT-ITS)}, 2019, pp. 1--7.

\bibitem{Nemade}
\BIBentryALTinterwordspacing
B.~Nemade, ``Automatic traffic surveillance using video tracking,'' \emph{Procedia Computer Science}, vol.~79, pp. 402--409, 2016, proceedings of International Conference on Communication, Computing and Virtualization (ICCCV) 2016. [Online]. Available: \url{https://www.sciencedirect.com/science/article/pii/S1877050916001836}
\BIBentrySTDinterwordspacing

\bibitem{sardar2019vehicle}
S.~Sardar, A.~K. Mishra, and M.~Z.~A. Khan, ``Vehicle detection and classification using lte-commsense,'' \emph{IET Radar, Sonar \& Navigation}, vol.~13, no.~5, pp. 850--857, 2019.

\bibitem{jana2023validation}
S.~Jana, D.~K.~K. Reddy, A.~K. Mishra, and M.~Z.~A. Khan, ``Validation of a commsense based isac system using in-situ mmwave propagation model,'' in \emph{2023 26th International Symposium on Wireless Personal Multimedia Communications (WPMC)}.\hskip 1em plus 0.5em minus 0.4em\relax IEEE, 2023, pp. 266--271.

\bibitem{Lim}
H.-S. Lim, H.-M. Park, J.-E. Lee, Y.-H. Kim, and S.~Lee, ``Lane-by-lane traffic monitoring using 24.1 ghz fmcw radar system,'' \emph{IEEE Access}, vol.~9, pp. 14\,677--14\,687, 2021.

\bibitem{Fang}
J.~Fang, H.~Meng, H.~Zhang, and X.~Wang, ``A low-cost vehicle detection and classification system based on unmodulated continuous-wave radar,'' in \emph{2007 IEEE Intelligent Transportation Systems Conference}, 2007, pp. 715--720.

\bibitem{Nguyen}
V.~C. Nguyen, D.~K. Dinh, V.~A. Le, and V.~D. Nguyen, ``Length and speed detection using microwave motion sensor,'' in \emph{2014 International Conference on Advanced Technologies for Communications (ATC 2014)}, 2014, pp. 371--376.

\bibitem{HB100Datasheet}
\BIBentryALTinterwordspacing
Theorycircuit, ``Hb100 microwave sensor datasheet,'' Online, 2016. [Online]. Available: \url{https://theorycircuit.com/wp-content/uploads/2016/09/HB100\_Microwave\_Sensor\_datasheet.pdf}
\BIBentrySTDinterwordspacing

\bibitem{Devalal}
S.~Devalal and A.~Karthikeyan, ``Lora technology - an overview,'' in \emph{2018 Second International Conference on Electronics, Communication and Aerospace Technology (ICECA)}, 2018, pp. 284--290.

\end{thebibliography}

\end{document}